\documentclass[twocolumn,secnumarabic,amssymb, showpacs, nobibnotes, aps, prd]{revtex4}

\usepackage{graphicx}
\usepackage{dcolumn}
\usepackage{bm}
\usepackage{color} 
\begin{document}


\title{Evidence for zero-differential resistance states in electronic bilayers}

\author{G. M. Gusev,$^1$ S. Wiedmann,$^{2,3}$ O. E. Raichev,$^4$ A. K. Bakarov,$^5$ and J. C. Portal$^{2,3,6}$}
\affiliation{$^1$Instituto de F\'{\i}sica da Universidade de S\~ao
Paulo, CP 66318, S\~ao Paulo, SP,Brazil}
\affiliation{$^2$Laboratoire National des Champs Magn\'{e}tiques
Intenses, CNRS-UJF-UPS-INSA, 38042 Grenoble, France}
\affiliation{$^3$INSA Toulouse, 31077 Toulouse Cedex 4, France}
\affiliation{$^4$Institute of Semiconductor Physics, NAS of Ukraine,
Prospekt Nauki 41, 03028, Kiev, Ukraine} \affiliation{$^5$Institute
of Semiconductor Physics, Novosibirsk 630090, Russia}
\affiliation{$^6$Institut Universitaire de France, 75005 Paris,
France}

\date{\today}

\begin{abstract}
We observe zero-differential resistance states at low temperatures and moderate direct currents
in a bilayer electron system formed by a wide quantum well. Several regions of vanishing resistance
evolve from the inverted peaks of magneto-intersubband oscillations as the current increases. The
experiment, supported by a theoretical analysis, suggests that the origin of this phenomenon is
based on instability of homogeneous current flow under conditions of negative differential
resistivity which leads to formation of current domains in our sample, similar to the case of
single-layer systems.
\end{abstract}


\maketitle

Studies of non-linear transport in high-quality two-dimensional electron system (2DES)
have revealed many interesting phenomena which occur in a perpendicular magnetic field at
large filling factors. In the presence of AC excitation by microwaves, there exist
microwave-induced resistance oscillations (MIROs) \cite{1} which obey the periodicity
$\omega/\omega_{c}$, where $\omega$ and $\omega_{c}$ are the radiation frequency and the cyclotron
frequency, respectively. The minima of these oscillations evolve into zero-resistance
states (ZRS) for high electron mobility and elevated microwave power \cite{2}. The MIROs
have been found also in bilayer and trilayer electron systems \cite{3}, where they
interfere with magneto-intersubband (MIS) oscillations \cite{4} because of the presence
of more than one populated subband. Recently, it has been demonstrated \cite{5} that ZRS exist
in bilayer systems despite of additional intersubband scattering. Stimulated by experimental
findings, theorists have proposed several microscopic mechanisms which reasonably explain
non-linear transport caused by microwave excitation \cite{6,7,8,9}.

A direct current (DC) excitation of high-quality 2DES leads to another group of non-linear
phenomena caused by Landau quantization and, therefore, distinct from electron heating effects
observed under similar conditions in samples with lower mobilities. The Hall-field induced resistance
oscillations (HIROs) are found in numerous experiments \cite{10,11,12,13} and are explained
\cite{10,14} in terms of large-angle elastic scattering between Landau levels (LLs) tilted
by the Hall field. Further, even a moderate DC causes a considerable decrease of
the resistance \cite{15}, which in high-quality samples may lead to the zero differential
resistance phenomenon. The zero-differential resistance states (ZdRS) found in
single-layer systems emerge either from the inverted maxima of Shubnikov-de
Haas (SdH) oscillations at relatively high magnetic fields \cite{16} or from a minimum of HIROs
\cite{17}. In the second case ZdRS appears at low magnetic fields, before the onset of
SdH oscillations, and extends over a continuous range of fields. The two seemingly
different regimes, however, are explained within the same model assuming formation of
current domains when the negative resistance conditions are reached and homogeneous current
picture becomes unstable \cite{18}. In order to learn more about the origin of ZdRS,
clear understanding of the domain model is required. In this connection, studies of the
systems which differ from standard 2DES are of crucial interest.

In this Rapid Communication, we report observation of ZdRS in a high-quality bilayer electron
system formed by a wide quantum well (WQW). Owing to scattering of electrons between two
populated 2D subbands, ZdRS in bilayer electron systems evolve from \textit{inverted MIS oscillations}
for relatively low DC and obey MIS oscillation periodicity. In contrast
to both regimes of ZdRS in single-layer systems \cite{16,17}, ZdRS occur in the
intermediate regime from overlapping to separated LLs (0.1~T$<B<$0.3~T). A theoretical consideration
of magnetoresistance in DC-driven electronic bilayers shows that negative differential
resistance (NDR) can be reached in our system, thereby supporting the domain model.

Our experiments are carried out on high-quality WQWs with a well width of $w$=45~nm, high
electron density $n_{s}\simeq 9.1 \times 10^{11}$~cm$^{-2}$ and a mobility of $\mu~\simeq 1.9 \times
10^{6}$~cm$^{2}$/V s after a brief illumination with a red light-emitting diode. Several samples
in Hall bar geometry (length $L=500$ $\mu$m and width $W=200$ $\mu$m) have been
studied, while we focus here on two samples (A and B, with a slightly higher electron density
for sample B). Measurements have been performed at mK temperatures in a dilution refridgerator
(base temperature 50~mK) (sample A) and up to 4.2~K in a cryostate with a variable
temperature insert (sample B). We record longitudinal resistance using a current of 0.5~$\mu$A
at a frequency of 13~Hz. Direct current $I_{DC}$ was applied simultaneously
through the same current leads to measure the differential resistance $r_{xx} = dV_{xx}/dI_{DC}$.

\begin{figure}[ht]
\includegraphics[width=9cm]{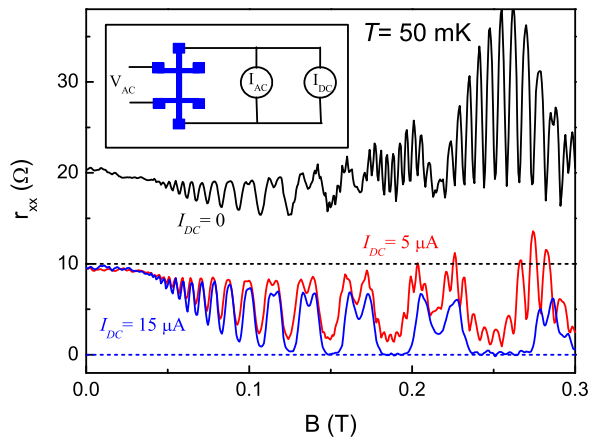}
\caption{\label{fig1} (Color online) Differential magnetoresistance $r_{xx}$ as a function of
the magnetic field at $I_{DC}$=5~$\mu$A and $I_{DC}$=15~$\mu$A for sample A. The top trace, which
is shifted up for clarity, shows magnetoresistance ($I_{DC}$=0) which exhibits MIS oscillations
together with SdH oscillations. Experimental setup is shown in the inset.}
\end{figure}

In Fig. \ref{fig1} we show differential resistance $r_{xx}$ as a
function of the magnetic field for $I_{DC}$=0, $I_{DC}$=5~$\mu$A and
$I_{DC}$=15~$\mu$A. Dark magnetoresistance exhibits well-developed
MIS oscillations and confirms the existence of two populated
subbands \cite{4}. The inversion of MIS oscillations for $B <
0.17$~T is caused by the alternating current of 0.5~$\mu$A. The position of
the inversion field is in a reasonable agreement with our
theoretical estimates \cite{19}. At this low temperature, we also
observe SdH oscillations which are superimposed on MIS oscillations
for $B>0.15$~T. If we apply a constant DC, the MIS
oscillations are inverted in the whole range of magnetic fields for
$I_{DC}$=5~$\mu$A at a temperature of 50~mK. For $I_{DC}$=15~$\mu$A,
we observe ZdRS for $B > 0.14$~T which are developed from the inverted MIS
oscillations. Having a closer look at magnetic fields 0.1~T$<B<$0.3~T, we see that
the maxima between ZdRS are split, which also occurs when applying a
strong AC or DC excitation separately \cite{19} and can be explained by
the influence of Landau quantization on inelastic scattering of
electrons in two-subband systems (see also Fig. \ref{fig4}).

As concerns temperature dependence, we have to apply a higher DC to maintain ZdRS at
higher temperature, see Fig \ref{fig4}(a) later. The reason for this is rooted in the fact that
the nonlinear response to a DC bias weakens with increasing temperature due to an increase
in electron-electron scattering \cite{9,14}.

\begin{figure}[ht]
\includegraphics[width=9cm]{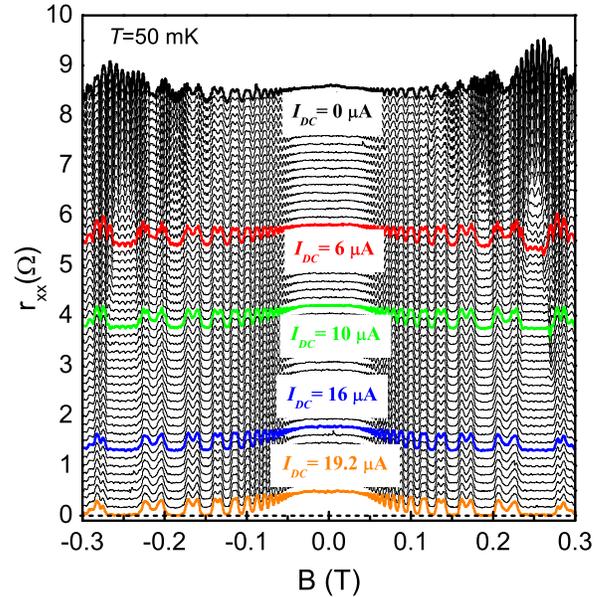}
\caption{\label{fig2} (Color online) Evolution of differential magnetoresistance $r_{xx}$ with
increasing DC bias from $I_{DC}$=0 to $I_{DC}$=19.2~$\mu$A (step $\Delta I_{DC}$=0.4~$\mu$A).
For the highest current, we find four ZdRS for $B>0.12$~T.}
\end{figure}

In order to give a more detailed overview of the evolution from MIS oscillations ($I_{DC}$=0)
to the regime where inversion of all MIS peaks takes place until we reach zero-differential
resistance, we plot in Fig. \ref{fig2} $r_{xx}$ as a function of the magnetic field starting from
$I_{DC}$=0 (top trace) to $I_{DC}$=19.2~$\mu$A (bottom trace) at a constant temperature of 50~mK for
both negative and positive magnetic field orientation in steps of $\Delta I_{DC}=0.4$ $\mu$A. Starting
from $I_{DC}$=16~$\mu$A the widths of ZdRS remain constant. A further increase in $I_{DC}$ leads to
heating of the electron gas in the sample, therefore our studies at $T$=50~mK are limited to a maximal
current of $I^{max}_{DC}$=20~$\mu$A.

\begin{figure}[ht]
\includegraphics[width=9cm]{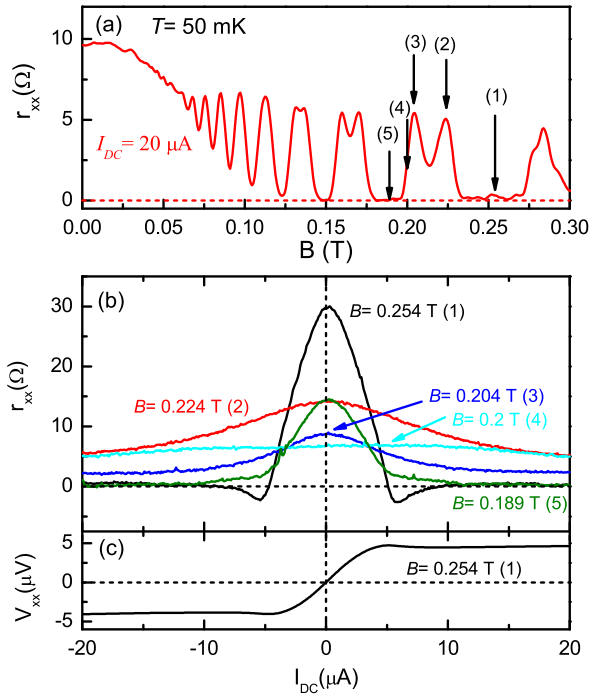}
\caption{\label{fig3} (Color online) (a) Differential resistance $r_{xx}$ for $I_{DC}$=20~$\mu$A.
(b) Dependence of $r_{xx}$ on the DC at several chosen magnetic fields as labeled at a temperature
of 50~mK. (c) Corresponding voltage $V_{xx}$ at $B$=0.254~T.}
\end{figure}

\begin{figure}[ht]
\includegraphics[width=9cm]{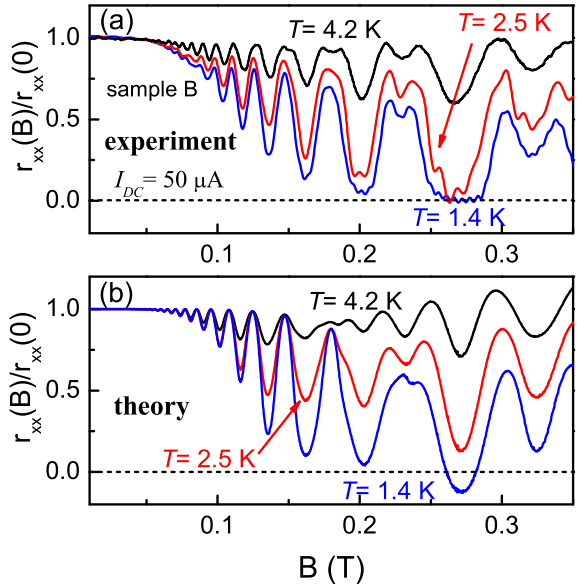}
\caption{\label{fig4} (Color online) (a) Differential magnetoresistance $r_{xx}$ in experiment and
(b) theory for three chosen temperatures (sample B).}
\end{figure}

Since ZdRS in electronic bilayers occur neither at nearly discrete values of $B$ \cite{16}
nor in a wide region of magnetic fields \cite{17}, and are separated by maxima in
$r_{xx}$, it is important to illustrate this phenomenon by sweeping $I_{DC}$ at a
constant magnetic field, see Fig. \ref{fig3}. We fix the magnetic field at several positions
(split maxima and ZdRS in $r_{xx}$), mark them by arrows with numbers (1) to (5) in Fig. \ref{fig3}(a),
and plot differential magnetoresistance $r_{xx}(I_{DC})$ in Fig. \ref{fig3}(b). For $B=0.254$~T and $B=0.189$~T,
we are situated in the center of ZdRS. Both ZdRS are developed at $I_{DC}>10$~$\mu$A.
For $B$=0.254~T we find small kinks in $r_{xx}$ at $\pm$5.8~$\mu$A before $r_{xx}$ is
stabilized near zero. Beyond ZdRS, we show similar plots for the two maxima in $r_{xx}$ at
$B=0.204$~T and $B=0.224$~T which occur between the ZdRS, and in the transition from these
maxima to ZdRS at $B=0.2$~T. The longitudinal voltage $V_{xx}$ as a function of $I_{DC}$ is
plotted in Fig. \ref{fig3}(c) for the ZdRS around $B$=0.25~T. This trace is obtained by
integrating the data at $B=0.254$~T and shows a typical voltage-current characteristic
with saturation of $V_{xx}$ corresponding to the onset of ZdRS.

We now analyze and discuss our experimental observation. First, we point out that the occurrence
of ZdRS in electronic bilayers is distinct from the case of single-layer systems where ZdRS emerge
in a wide range of magnetic fields well below the onset of SdH oscillations. In a two-subband system
several ZdRS emerge from the peaks of inverted MIS oscillations and follow the $1/B$ periodicity
of these oscillation determined by the subband separation $\Delta$. In other words, each ZdRS
appears around the magnetic fields given by the set of cyclotron energies $\hbar \omega_{c} =\Delta/k$,
where $k$ is an integer. Next, the other important properties of ZdRS, such as the dependence
of $r_{xx}$ on the DC and the voltage-current characteristics, are very similar to those
reported for single-layer systems \cite{16,17}.

From the theoretical point of view, it is quite possible that the differential dissipative resistance
$r_{xx}= dV_{xx}/dI_{DC}$ (or even absolute resistance $R_{xx}$) in high-mobility samples approaches
zero with increasing current, since an application of the current leads to a considerable decrease in
the resistivity. Such a decrease, observed both in single-layer \cite{15} and double-layer
\cite{19} systems, is reasonably explained by the theory of Ref. \cite{9} based on a modification
of the electron distribution function due to current-induced diffusion of electrons in the energy
space in the presence of Landau quantization. This non-equilibrium distribution function is
stabilized by inelastic electron-electron collisions described through the inelastic
scattering time $\tau_{in}$ (inelastic mechanism). A more rigorous theory, which includes
another (displacement) mechanism of nonlinear response and remains valid at higher currents,
has been developed in Ref. \cite{14}. To investigate the possibility of NDR for our
samples and to explain other characteristic features of the observed magnetoresistance, we
have generalized the theory of Ref. \cite{14} beyond the regime of overlapping Landau levels
and for the case of multisubband occupation. Below we present the results valid both for
single-subband systems and for the systems with two closely spaced subbands (our samples).
In the regime of classically strong magnetic fields, the magnetoresistance of the system of
electrons interacting with the static potential of impurities is found from the following relation
between the density of the applied current $j=I_{DC}/W$ and longitudinal electric field $E_{||}=V_{xx}/L$:
\begin{equation}
E_{||}=j \rho_0 \biggl[1-\sum_{k (k\neq 0)} a_k^2 \gamma''_k - \sum_{kk'} a_k a_{k-k'} A_{k'}
\left( \gamma'_k - \gamma'_{k-k'} \right) \biggr],
\end{equation}
where $\rho_0=m/e^2n_s \tau_{tr}$ is the classical Drude resistivity ($e$ and $m$ are
the electron charge and the effective mass). The contribution of SdH
oscillations is neglected. The sums are taken over the cyclotron harmonics, $k$ are integers,
$a_k$ are the coefficients of expansion of the density of electron states in such harmonics,
$D(\varepsilon)=\sum_k a_k \exp(i k \frac{2 \pi \varepsilon}{\hbar \omega_c} )$, where
$D(\varepsilon)$ is the density of states in units of $m/\pi \hbar^2$.
For two-subband systems, one should use $D(\varepsilon)=(D_{1\varepsilon}+D_{2\varepsilon})/2$,
where $D_{i\varepsilon}$ is the density of states in the subband $i$. In both cases $a_k$ are
taken real, which is attainable by a proper choice of the reference point for energy. The
coefficients $A_{k'}$, which describe harmonics of non-equilibrium oscillating correction to
the distribution function, are found from the linear system of equations
$\sum_{k'}C_{kk'}A_{k'}=a_k \gamma'_k$ with
\begin{equation}
C_{kk'}= \frac{\tau_{tr}}{\tau_{in}}[a^3_{k-k'}+a^2_{k}(a_{k+k'}-2a_{k-k'})]
+ a_{k-k'}(\gamma_{k-k'}-\gamma_{k}).
\end{equation}
The first part of the matrix $C_{kk'}$ is determined by taking into account the Landau quantization
in the linearized electron-electron collision integral (see details in Ref. \cite{19}).
The coefficients $\gamma_k$ describe the effect of the current and are defined by
averaging over the scattering angle $\theta$: $\gamma_k \equiv \gamma(\zeta k)=
\overline{ J_0[2 \zeta k \sin(\theta/2)] \tau_{tr}/\tau(\theta) }$, where $J_0(x)$ is the
Bessel function, $\zeta=\sqrt{4 \pi^3 j^2/e^2 n_s \omega_c^2}$ (for single-subband
systems one should multiply $\zeta$ by $\sqrt{2}$), and $1/\tau(\theta)=(m/\hbar^3)
w[2k_F\sin(\theta/2)]$ is the angular-dependent scattering rate ($k_F$ is the Fermi
wavenumber, $w(q)$ is the correlator of random scattering potential). The
transport time $\tau_{tr}$ is given in the usual way, $1/\tau_{tr}=\overline{ (1-\cos
\theta)/\tau(\theta) }$. Finally, $\gamma'_k$ and $\gamma''_k$ denote, respectively, the
first and the second derivatives of the function $\gamma(\zeta k)$ over its argument.

To carry out the calculations, we numerically computed the density of states in the self-consistent
Born approximation, using the quantum lifetime of electrons $\tau_0=6.6$ ps determined from linear
low-temperature magnetoresistance measurements. Next, we applied the theoretical \cite{9} estimate
$\tau_{in} \simeq \hbar \varepsilon_F/T_e^2$, where $\varepsilon_F$ is the Fermi energy and $T_e$
is the electron temperature. Heating of electron gas by the current is taken into account by considering
energy loss due to electron-phonon interaction \cite{3,20} (for $I_{DC}=50$ $\mu$A the heating is
appreciable only at low temperatures, at $T=1.4$~K we find $T_e \simeq 1.6$~K). To describe elastic
collisions, we used the model of long-range scattering potential, $w(q) \propto \exp(-l_cq)$
(where $l_c \gg 1/k_F$ is the correlation length), which produces $\tau_{tr}=\tau_0(1+\chi)/\chi$
with $\chi=1/(k_F l_c)^2$ and $\gamma(\zeta) \simeq (\tau_{tr}/\tau_0)/\sqrt{1+\chi \zeta^2}$ \cite{14}.

In Fig. \ref{fig4} the calculated differential magnetoresistance
plots are compared with experimental data recorded on sample B with a subband separation energy
$\Delta =1.4$ meV. Notice that this separation is slightly larger than in sample A, so the
positions of inverted MIS peaks (and of the corresponding ZdRS) are slightly shifted towards
higher magnetic fields with respect to sample A. Applying $I_{DC}=50$~$\mu$A,
we observe ZdRS around $B=0.27$~T at low temperatures (see the plot for $T=1.4$ K) and its
disappearance at $T > 2.5$~K. The calculation shows a similar behavior with a NDR region around
$B=0.27$~T. The other important feature of magnetoresistance, the peak splitting leading to
minima at 0.23~T and 0.32~T, is also reproduced in the calculations, see Ref. \cite{19} for a
detailed description of this phenomenon.
Therefore, a direct calculation of magnetoresistance demonstrates that NDR is attainable for
our samples. In the NDR regions, according to Ref. \cite{18}, the homogeneous flow of the current
becomes unstable and the system exhibits a transition to current domains. The simplest two-domain
structure discussed in Refs. \cite{16,17} can as well describe ZdRS in our experiment.

To summarize, we have found evidence for zero-differential resistance in a bilayer (two-subband)
electron system where ZdRS develop from inverted magento-intersubband oscillations under a relatively
small direct current. This experimental result, together with a theoretical consideration, suggests
that the ZdRS occurs as a result of current instability under negative differential resistance
conditions. The domain model proposed for explanation of ZdRS in single-subband 2DEG \cite{16,17}
very likely applies to multisubband systems.

\emph{Note added in proof.} We have recently become aware of related
experimental work on a similar system, by Ref. 21.

We acknowledge support from COFECUB-USP (project number U$_{c}$ Ph
109/08), FAPESP and CNPq (Brazilian agencies).

\end{document}